\documentclass[%
 reprint,
 amsmath,amssymb,
 aps,
]{revtex4-2}

\usepackage{graphicx}
\usepackage{dcolumn}
\usepackage{bm}
\usepackage{amsmath}
\usepackage{siunitx}

\begin{document}

\preprint{APS/123-QED}

\title{The ability of Lisa, Taiji, and their networks to detect the stochastic gravitational wave background generated by Cosmic Strings}

\author{Bo-Rui Wang}
 \affiliation{College of Physics, Chongqing University, Chongqing 401331, China}
\author{Jin Li}%
 \email{ cqujinli1983@cqu.edu.cn}
\affiliation{%
 College of Physics, Chongqing University, Chongqing 401331, China\\
 Department of Physics and Chongqing Key Laboratory for Strongly Coupled Physics, Chongqing University, Chongqing 401331, China
}%

\date{\today}

\begin{abstract}
The cosmic string contributes to our understanding and revelation of the fundamental structure and evolutionary patterns of the universe, unifying our knowledge of the cosmos and unveiling new physical laws and phenomena. Therefore, we anticipate the detection of Stochastic Gravitational Wave Background (SGWB) signals generated by cosmic strings in space-based detectors. We have analyzed the detection capabilities of individual space-based detectors, Lisa and Taiji, as well as the joint space-based detector network, Lisa-Taiji, for SGWB signals produced by cosmic strings, taking into account other astronomical noise sources. The results indicate that the Lisa-Taiji network exhibits superior capabilities in detecting SGWB signals generated by cosmic strings and can provide strong evidence. The Lisa-Taiji network can achieve an uncertainty estimation of $\Delta G\mu/G\mu<0.5$ for cosmic string tension $G\mu\sim4\times10^{-17}$, and can provide evidence for the presence of SGWB signals generated by cosmic strings at $G\mu\sim10^{-17}$, and strong evidence at $G\mu\sim10^{-16}$. Even in the presence of only SGWB signals, it can achieve a relative uncertainty of $\Delta G\mu/G\mu<0.5$ for cosmic string tension $G\mu<10^{-18}$, and provide strong evidence at $G\mu\sim10^{-17}$.
\end{abstract}

\maketitle


\section{\label{sec:level1}INTRODUCTION}

Using space-based laser interferometers to~explore gravitational wave sources is currently a hot research topic.~As space detectors, The Laser Interferometer Space Antenna (Lisa)\cite{1} and Taiji\cite{2} are sensitive to gravitational waves in the millihertz frequency range and can explore gravitational wave signals emitted by independent sources from astronomy and cosmology, as well as Stochastic Gravitational wave background (SGWB) signals generated by a large number of independent sources. Exploring cosmological SGWB is of great significance for studying the early behavior of the universe and important for testing the universe models. The SGWB may come from many different processes in the early universe such as phase transitions\cite{3,4,5,6}  inflation models\cite{7,8}, and cosmic strings\cite{9,10,11}. The corresponding frequency of Gravitational Wave (GW) signal is in $(10^{-18}-10^{10} Hz)$\cite{12,13}.

In practical detection, any cosmological gravitational wave signal will be mixed with other foreground and background noise. Apart from cosmological SGWB, there is astronomical SGWB originated from the superposition of gravitational waves generated by a large number of celestial bodies. In this paper, we need to separate the cosmological SGWB from the noise to understand the behavior of the universe at that time. Here we are concerned with the SGWB signal generated by cosmic strings and assumes that the mixed foreground noise consists of two parts: one is the gravitational wave background (GWB) model \cite{
14,15}, generated by binary black holes (BBH) / binary neutron stars (BNS) based on observations of the stellar mass black hole from LIGO and Virgo, and the other is the SGWB from unresolved White Dwarf Binaries in our galaxy, which is observed as a modulated waveform due to Lisa's orbital motion \cite{16,17}.

As one of the most prospective approach for detecting SGWB, space detectors need to understand their sensitivity to cosmological string GW signals and their ability to separate them from confusing noise. Several teams have now conducted detailed researches on the capabilities of Lisa\cite{18,19}. The result show that Lisa has good identification and estimation capabilities for SGWB signals and their associated parameters generated by first-order transitions and cosmic strings in the presence of contained noise. Therefore it is also important to understand the corresponding capabilities of Taiji as a Lisa-like detector. Moreover, a single detector is unable to locate space sources very well\cite{20}. For this point, the proposed joint Lisa-Taiji observation can be expected to significantly improve the accuracy of source location\cite{21,22} and detectability\cite{23}. Consequently, studying the sensitivity of joint space networks to detect SGWB from cosmological strings and their ability to separate it from confusing noise is also attractive. In this paper, the structure of joint space network is constructed by the three Taiji detector orbit designs mentioned in\cite{24}.
Cosmic strings are one-dimensional topological defects\cite{25}, which may be produced by spontaneous symmetry breaking after a phase transition in the early universe and are expected to exist throughout cosmic history. Some results have shown that the Lisa detector can detect cosmic strings with tension $G\mu\gtrsim O(10^{-17})$ under any cosmic string model\cite{26,27,28,29}. In previous work, we also used Taiji and Lisa-Taiji joint networks to detect cosmic strings SGWB in Model 2\cite{30} and the results showed that the joint networks are also able to detect cosmic strings with tension $G\mu\gtrsim O(10^{-17})$\cite{31}. Therefore, we hope to further understand the detectability of different space-based millihertz GW detectors to cosmological string SGWB with confusing foregrounds, such as, the superposition of GWs from double white dwarfs and BBH/BNS (see Sec.\ref{IV}), the use Fisher matrix for parameter estimation and Deviance Information Criterion (DIC) method to more intuitively demonstrate the detectability of the detectors for observing the SGWB in different cosmic string models.

The structure of this paper is as follows. In Sec.\ref{II}, we describe the gravitational wave background from cosmic strings, double white dwarfs, and stellar-mass black hole in inspiral stage based on LIGO and Virgo observation. In Sec.\ref{III}, we discuss the noise model of detectors, the sensitivity curves of joint networks and single detectors to cosmic strings, and analyze the possibility of identifying cosmological string SGWB from foreground and background noise. In Sec.\ref{IV}, we introduce how to use the Fisher matrix and Bayesian factor DIC to calculate the results of cosmic string parameter estimation through the cosmic string SGWB detection. Finally, we summarize the results and give conclusions in Sec.\ref{V}.

\section{\label{II}COMPOSITION OF SGWB SOURCES IN OUR WORK }

The stochastic gravitational wave background (SGWB) concern in this paper consists of three parts: the cosmic string stochastic gravitational wave signal $\Omega_{GW}$, the double white dwarf foreground $\Omega_{DWD}$, and the GW foreground $\Omega_{astro}$ generated by BBH/BNS based on observations of the inspiral stellar mass black hole in LIGO and Virgo. Among them, the stochastic gravitational wave generated by cosmic strings $\Omega_{GW}$ is the signal that we hope to identify, and the other two parts $(\Omega_{DWD},\Omega_{astro})$ are considered as confusing foreground noise.

\subsection{\label{II.1}Gravitational Wave From Cosmic strings}

The stochastic gravitational wave background of cosmic strings is a non-coherent superposition of gravitational waves emitted by oscillating cosmic string loops. There has been extensive research on the stochastic gravitational wave background generated by cosmic strings\cite{10,27,32,33,34,35,36,37,38,39,40,41,42,43,44,45,46,47,48,49,50,51}, which includes two analytical methods and three cosmic string models commonly used for calculating cosmic strings.

We use the template mentioned in\cite{30,32} to represent the gravitational wave of cosmic strings, and there are exact analytical approximation formulas for Model 1 and Model 2. The GW we studied is a function of the cosmic string tensor $G\mu$, which characterizes the size of the loop with a free constant $\alpha$ that we consider as a constant value, i.e., $\alpha=0.1$, and we define the total power of cosmic string emission as $\Gamma=50$. For the gravitational wave signals of Model 1 and Model 2 there are a total of three periods of the gravitational wave contribution of the cosmic string loop: loops formed and decayed during the radiation period, loops formed during the radiation period and decayed during the matter period, and loops formed during the matter period.

For loops formed and decayed in the radiation region, the form of stochastic gravitational wave background is given by
\begin{eqnarray}
	\Omega_{\mathrm{GW}}^r(f)=\frac{128}{9}\pi A_r\Omega_r\frac{G\mu}{\epsilon_r}\left[\left(\frac{f(1+\epsilon_r)}{\frac{B_r\Omega_m}{\Omega_r}+f}\right)^{\frac{3}{2}}-1\right],
\end{eqnarray}
where $\epsilon_r=\frac{\alpha}{\Gamma G\mu}$,$\Omega_r$is radiation energy density ratio, $A_r=0.54$, and
\begin{eqnarray}
	B_r=\frac{2H_0\Omega_r^{\frac{1}{2}}}{\nu_r\Gamma G\mu},
\end{eqnarray}
where $\nu_r=\frac{1}{2}$.
For loops formed in the radiation region and decayed in the matter region, their contribution to SGWB has the following form
\begin{widetext}
	\begin{equation}
		\Omega_{\mathrm{GW}}^{rm}(f)=32\sqrt{3}\pi(\Omega_{m}\Omega_{r})^{\frac{3}{4}}H_{0}\frac{A_{r}}{\Gamma}\frac{(\epsilon_{r}+1)^{\frac{3}{2}}}{f^{\frac{1}{2}\epsilon_{r}}}
		\left\{\frac{\left(\frac{\Omega_m}{\Omega_r}\right)^{\frac{1}{4}}}{\left(B_m\left(\frac{\Omega_m}{\Omega_r}\right)^{\frac{1}{2}}+f\right)^{\frac{1}{2}}}\left[2+\frac{f}{B_m\left(\frac{\Omega_m}{\Omega_r}\right)^{\frac{1}{2}}+f}\right]-\frac{1}{(B_m+f)^{\frac{1}{2}}}\left[2+\frac{f}{B_m+f}\right]\right\},\label{eq3}
	\end{equation}
\end{widetext}
where $\Omega_{m}$ is matter energy density ratio, and
\begin{eqnarray}
	B_m=\frac{2H_0\Omega_m^{\frac12}}{\nu_m\Gamma G\mu},
\end{eqnarray}
where $\nu_m=2/3$.
The contribution of loops generated in the matter period to the SGWB generation by cosmic strings is given by
\begin{widetext}
	\begin{equation}
	\Omega_{\mathrm{GW}}^m(f)=54\pi H_0\Omega_m^{\frac32}\frac{A_m}\Gamma\frac{\epsilon_m+1}{\epsilon_m}\frac{B_m}f\Bigl\{\frac{2B_m+f}{B_m(B_m+f)}-\frac1f\frac{2\epsilon_m+1}{\epsilon_m(\epsilon_m+1)}+\frac2f\log\Bigl(\frac{\epsilon_m+1}{\epsilon_m}\frac{B_m}{B_m+f}\Bigr)\Bigr\}.
	\end{equation}
\end{widetext}

Therefore, for Model 1 and Model 2, the SGWB generated by cosmic strings can be well approximated as
\begin{eqnarray}
	\Omega_{\mathrm{GW}}(f,M_{1,2})=\Omega_{\mathrm{GW}}^{r}(f)+\Omega_{\mathrm{GW}}^{rm}(f)+\Omega_{\mathrm{GW}}^{m}(f),\label{eq6}
\end{eqnarray}
which can provide a good approximation for loops with $\alpha\geq\Gamma G\mu $\cite{30,32}. For some small loops, i.e., those whose size cannot support their survival from radiation to matter-dominated era, the SGWB in Eq. \eqref{eq3} will not be included in Eq. \eqref{eq6}\cite{30,51}.

For Model 3, we still use an analytical approximation model, which was summarized in\cite{34} firstly. The analytical approximation model we used comes from\cite{32,52}. Unlike Model 1 and Model 2, Model 3 includes two additional parts of loop contributions besides the three mentioned above which two extra contributions are small loops that exist and decayed during the radiation and matter periods. That is to say, under this model, we need to consider loops with length $\alpha\geq\Gamma G\mu$ and an extra population of small loops with invariant lengths smaller than $\Gamma G\mu t$\cite{32}.Therefore, for Model 3, the SGWB generated by cosmic strings includes five parts of contributions.

The form of gravitational wave produced by loops during the radiation period is given by
\begin{eqnarray}
	\Omega_{\mathrm{GW}}^r=\frac{64\pi C_r\Omega_r}{3\Gamma(2-2\chi_r)}(\Gamma G\mu)^{2\chi_r}\left(1+\frac{4H_r\big(1+z_{\mathrm{eq}}\big)}{f\Gamma G\mu}\right)^{2\chi_r-2},
\end{eqnarray}
where $C_r=0.08$,$\chi_r=0.2$,$H_r$is the Hubble function during radiation domination period, and $z_{eq}$ is the redshift when the matter and radiation energy densities are equal, here $z_{eq}=3400$\cite{53}. For cosmic string loops formed during radiation domination period but existed during matter domination period, their contribution to SGWB is given by
\begin{widetext}
	\begin{equation}
		\Omega_{\mathrm{GW}}^{rm}=\frac{54\pi C_{r}H_{m}\Omega_{m}}{\Gamma f(\Gamma G\mu)^{1-2\chi_{r}}}(1+z_{\mathrm{eq}})^{\frac{3(2\chi_{r}-1)}{2}}\left[\frac{x^{2-6\chi_{r}}}{2-6\chi_{r}}{}_{2}F_{1}\left(3-2\chi_{r},2-6\chi_{r};3-6\chi_{r};-\frac{3H_{m}x}{f\Gamma G\mu}\right)\right]_{1}^{\sqrt{1+z_{eq}}}.
	\end{equation}
\end{widetext}

The square bracket's superscript and subscript represent the upper and lower limits of integration.  ${}_2F_1(a,b;c;d)$ is the Gaussian hypergeometric function
\begin{subequations}
	\begin{eqnarray}
		_2F_{1}(a,b;c,d)=\sum_{n=0}^{\infty}\frac{(a)n(b)n}{(c)n}\frac{d^{n}}{n!},
	\end{eqnarray}
	\begin{equation}
		(a)_n=a(a+1)(a+2)\cdots(a+n-1)=\frac{\Gamma(a+n)}{\Gamma(a)},
	\end{equation}
 \end{subequations}
 here $(a)_n$ is Pochhammer symbol and $\Gamma(x)$ is Gamma Function
 . $H_m$ is the Hubble function during matter domination period, and for loops formed during matter domination period, their contribution to SGWB is given by
 \begin{widetext}
 	\begin{equation}
 		\left.\Omega_{\mathrm{GW}}^m=\frac{2\times3^{2\chi_m}\pi C_m\Omega_m}{H_m^{2-2\chi_m}\Gamma f^{2\chi_m-2}}(\Gamma G\mu)^2\left[\frac{x^{2\chi_m-4}}{2\chi_m-4}\right._2F_1\left(3-2\chi_m4-2\chi_m;5-2\chi_m;-\frac{f\Gamma G\mu}{3H_mx}\right)\right]_1^{\sqrt{1+z_{\mathbf{eq}}}},
 	\end{equation}
 \end{widetext}
 where $\chi_m=0.295$, $C_m=0.015$. In addition to the stochastic gravitational wave signals generated by the three parts of loops mentioned above, two additional sets of small loops still have a significant contribution to this model. However, the sizes of these two groups of string loops are too small to survive from the radiation-dominated period to the matter-dominated period. Therefore the population of additional small loops has only two contributions as follows: (1) Their contribution during the radiation-dominated period is given by
 \begin{widetext}
 	\begin{eqnarray}
 	\Omega_{\mathrm{GW}}^{r, \mathrm{epsl}}=&&\frac{64 \pi C_{r} \Omega_{r}\left(1 / 2-2 \chi_{r}\right)}{3\left(1-2 \chi_{r}\right)\left(2-2 \chi_{r}\right)} G \mu \gamma_{c}^{2 \chi_{r}-1} \\
 	\times&&\left\{\begin{array}{ll}0 & \text { if } f<4\left(1+z_{\mathrm{eq}}\right) H_{r}(\Gamma G \mu)^{-1}  \\{\left[4\left(1+z_{\mathrm{eq}}\right) H_{r} /\left(\gamma_{c} f\right)\right]^{2 \chi_{r}-1}-\left(\Gamma G \mu / \gamma_{c}\right)^{2 \chi_{r}-1}} & \text { if } f<4\left(1+z_{\mathrm{eq}}\right) H_{r} \gamma_{c}^{-1} \\\left(2-2 \chi_{r}\right)-4\left(1+z_{\mathrm{eq}}\right) H_{r}\left(1-2 \chi_{r}\right) /\left(\gamma_{c} f\right)-\left(\Gamma G \mu / \gamma_{c}\right)^{2 \chi_{r}-1} & \text { if } f>4\left(1+z_{\mathrm{eq}}\right) H_{r} \gamma_{c}^{-1}\end{array}\right.\nonumber.
 	\end{eqnarray}
 \end{widetext}
 
 (2) For small loops during the matter-dominated period, their contribution to the stochastic gravitational wave background generated by cosmic strings also has a piecewise function form
 \begin{widetext}
 	\begin{eqnarray}
 		\Omega_\mathrm{GW}^{m,\mathrm{epsl}}=&&\frac{54\pi C_mH_m\Omega_m(1-2\chi_m)}{(3-2\chi_m)(2-2\chi_m)f}G\mu\gamma_c^{2\chi_m-2}\left(\frac{3H_m}{\gamma_cf}\right)\\
 		\times&&\begin{cases}0&\mathrm{if~}f<4H_m(\Gamma G\mu)^{-1}\\\left[\frac{3H_m}{\gamma_c}\right]^{2\chi_m-3}\left\{1-[3H_m/(\Gamma G\mu f)]^{3-2\chi_m}\right\}&\mathrm{if~}f<4H_m\sqrt{1+z_\mathrm{eq}}(\Gamma G\mu)^{-1}\\\left[\frac{3H_m}{\gamma_cf}\right]^{2\chi_m-3}\left[1-(1+z_\mathrm{eq})^{-(3-2\chi_m)/2}\right]&\mathrm{if~}f<4H_m\gamma_c^{-1}\\(3-2\chi_m)f\gamma_c/(3H_m)+(2\chi_m-2)-\left(\frac{f\gamma_c}{3H_m\sqrt{1+z_\mathrm{eq}}}\right)^{3-2\chi_m}&\mathrm{if~}f<4H_m\sqrt{1+z_\mathrm{eq}}\gamma_c^{-1}\\\left[\frac{3H_m}{\gamma_cf}\right]^{-1}(3-2\chi_m)\left[1-(1+z_\mathrm{eq})^{-1/2}\right]&\mathrm{if~}f>4H_m\sqrt{1+z_\mathrm{eq}}\gamma_c^{-1}\end{cases}\nonumber.
 	\end{eqnarray}
 \end{widetext}

 Therefore, for Model 3 the form of the stochastic gravitational wave signal generated by cosmic strings should be
 \begin{eqnarray}
 	\Omega_{\mathrm{GW}}(f,M_3)&&=\Omega_{\mathrm{GW}}^{r}(f)+\Omega_{\mathrm{GW}}^{rm}(f)+\Omega_{\mathrm{GW}}^{m}(f)\nonumber\\
 	&&+\Omega_{\mathrm{GW}}^{r,\mathrm{epsl}}+\Omega_{\mathrm{GW}}^{m,\mathrm{epsl}}.
 \end{eqnarray}
 \subsection{Gravitational Wave Background}
 In actual detection, the data includes not only the gravitational wave signals generated by cosmic strings but also astronomical foreground noise. The foreground noise include the superposition of GW from double white dwarf	(DWD) and inspiralling BBH/BNS based on the observations of LIGO and Virgo\cite{14,18,57}.
 
 The model of GW from double white dwarf is a modulated signal based on the Lisa orbital motion, and its energy spectral density can be approximated by the broken power-law model proposed by Lambert et al.\cite{18,19}, which is given by
 \begin{eqnarray}
 	\Omega_{\mathrm{DWD}}(f)=\frac{A_1\left(\frac{f}{f_*}\right)^{\alpha_1}}{1+A_2\left(\frac{f}{f_*}\right)^{\alpha_2}},\label{eq.14}
 \end{eqnarray}
 it should be noted that for calculating the GW of double white dwarf, we use the model based on the modulation signal generated by Lisa for different detectors. Therefore, for this paper at any detector, the GW produced by DWD is treated as Eq.\eqref{eq.14}, that is, $f_*=c/2\pi L_*$, $L_*=2.5\times10^6km$. For the superposition of gravitational wave background produced by inspiralling BBHs/BNS observed by LIGO and Virgo, it can be modeled as a power-law function based on the observation results. This model is consistent with the one used in references [18,19], which is given by
 \begin{eqnarray}
 	\Omega_{astro}(f)=\Omega_{astro}\left(\frac{f}{f_*}\right)^{\alpha_{astro}},
 \end{eqnarray}
where $f_*=3mHZ$.

Therefore, the total energy spectrum related to gravitational waves discussed in this paper is as follows
\begin{eqnarray}
	\Omega_{\mathrm{tot}}(f)=\Omega_{astro}(f)+\Omega_{\mathrm{DWD}}(f)+\Omega_{\mathrm{GW}}(f,M_{x}).
\end{eqnarray}

Where $M_x (x=1,2,3)$ represents the three different gravitational wave models generated by cosmic string loops, namely Model 1, Model 2, and Model 3. For the GW from double white dwarf and the superposition of BBHs/BNS, we set the parameters as shown in Table~\ref{table1}.
\begin{table}[h]
	\caption{\label{table1}%
		Parameter values for astronomical foreground noise in data simulation.
	}
	\begin{ruledtabular}
		\begin{tabular}{cccc}
			\textrm{Parameter}&
			\textrm{Value}&
			\textrm{Parameter}&
			\textrm{Value}\\
			\colrule
			$A_1$ & $7.44\times10^{-14}$ & $A_2$ & $2.96\times10^{-7}$\\
			$\alpha_1$ & -1.98 & $\alpha_2$ & -2.6\\
			$\Omega_{astro}$ & $4.44\times10^{-12}$ & $\alpha_{astro}$ & 2/3\\
		\end{tabular}
	\end{ruledtabular}
\end{table}

\begin{figure}[h]
	\includegraphics[width=\columnwidth]{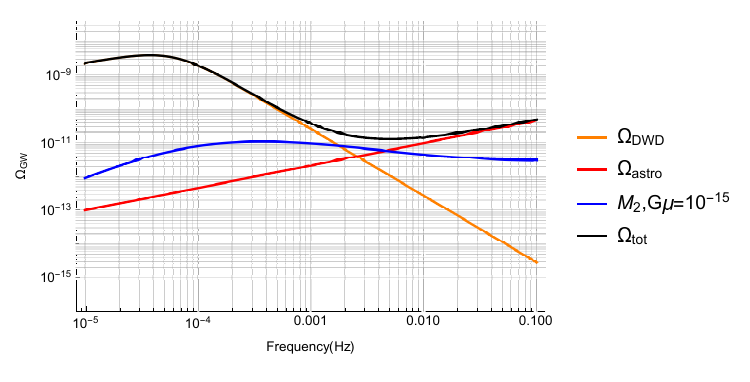}
	\caption{\label{figure1}The orange solid line represents the GW from double white dwarf, the red solid line represents the gravitational wave superposition background during the black hole inspiralling of stellar-mass black holes, the blue solid line represents the gravitational wave generated by cosmic strings with string tension $G\mu=10^{-15}$ in Model 2, and the black solid line represents the total energy spectrum of the gravitational waves generated by the three sources mentioned above.}
\end{figure}

The energy density spectra for the above gravitational wave sources are displayed in Figure~\ref{figure1}, including GWs from double white dwarf and inspiralling BBHs/BNS and cosmic string loops.
\section{\label{III}THE SPACE DETECTORS MODEL }
\subsection{Noise models for Lisa and Taiji}

Using the Time Delay Interferometry (TDI) technique, the laser frequency noise of Lisa and Taiji can be suppressed\cite{19,24,55,56,57,58,59,60,61,62,63,64,65,66,67,68,69,70,71,72}. For Lisa-like detectors, three suitable gravitational wave measurement channels, namely A, E, and T channels, can be constructed through the TDI technique. Assuming that the arm length of the Lisa-like detector is stable and the responses in these three channels are stable and uncorrelated, we consider only the responses in the A and E channels, as there is no response in the T channel and only instrument noise is present in this "null" channel\cite{18,19}. At the same time, since the effect of the T channel is not significant in the range of our concern frequency band\cite{31,56,73,74,75}, in the following calculations we only consider the responses of the A,E channels. The gravitational wave signal response in the A and E channels of the Lisa-like detector is given by\cite{75}
\begin{equation}
	R_A^i(f)=R_E^i(f)=\frac{9}{20}\mid W^i(f)\mid^2\left[1+\left(\frac{f}{\frac{4f_i}{3}}\right)^2\right]^{-1},
\end{equation}
where $i=Lisa,Taiji$, $W^{i}(f)=1-e^{-2if/f_{i}}$, and for the Lisa-like detector, $f_i=c/2\pi L_i$, with $L_{Lisa}=2.5\times10^6km$ and $L_{Taiji}=3\times10^6km$.

Based on the noise model given in the LISA Science Requirement Document\cite{19,76}, Lisa noise consists of acceleration and optical path disturbance noise. Similarly, for the Lisa-like detector Taiji, the model we use is similar to that of Lisa, which has the same acceleration noise and slightly different optical path noise\cite{1,2,24}.  For Lisa, its acceleration and optical path disturbance are given by
\begin{subequations}
	\begin{eqnarray}
		&&\sqrt{(\delta a_{Lisa})^2}=3\times10^{-15}\mathrm{m/s}^2,\\
		&&\sqrt{(\delta x_{Lisa})^2}=1.5\times10^{-11}\mathrm{m}.
	\end{eqnarray}
\end{subequations}

While for Taiji they are\cite{21}
\begin{subequations}
	\begin{eqnarray}
		&&\sqrt{(\delta a_{Taiji})^2}=3\times10^{-15}\mathrm{m/s}^2,\\
		&&\sqrt{(\delta x_{Taiji})^2}=8\times10^{-12}\mathrm{m}.
	\end{eqnarray}
	
\end{subequations}
The acceleration and optical path disturbance noise are
\begin{subequations}
	\begin{eqnarray}
		N_{acc}^i(f)=\frac{N_a^i}{(2\pi f)^4}\Bigg(1+\left(\frac{f_1}f\right)^2\Bigg)\nonumber\\
		=\frac{(\sqrt{(\delta a_i)^2}/L_i)^2}{(2\pi f)^4}\Bigg(1+\left(\frac{f_1}f\right)^2\Bigg)\mathrm{Hz}^{-1},
	\end{eqnarray}
	\begin{equation}
		N_{op}^i(f)=N_o^i=(\sqrt{(\delta x_i)^2}/L_i)^2\mathrm{Hz}^{-1},
	\end{equation}
\end{subequations}
where $i=Lisa,Taiji$, $f_1=0.4m$Hz. These noise models can be transformed into interferometer noise through
	\begin{equation}
		N_1^i(f)=\left[4N_{op}^i(f)+8[1+cos^2(f/f_i)]N_{acc}^i(f)\right]|W^i(f)|^2,
	\end{equation}
	\begin{equation}
		N_2^i(f)=-\bigl[2N_{op}^i(f)+8N_{acc}^i(f)\bigr]\cos(f/f_i)|W^i(f)|^2.
	\end{equation}
	
	We can construct the noise power spectral density of the A and E channels of the detector through
\begin{eqnarray}
	N_A^i(f)=N_E^i(f)=N_1^i(f)-N_2^i(f).
\end{eqnarray}

The noise spectral density formula for different channels can be constructed from the noise power spectral density and response function
\begin{eqnarray}
	S_{A}^{i}(f)=S_{E}^{i}(f)=\frac{N_{A,E}^{i}}{R_{A,E}^{\mathrm{i}}(f)}\label{eq24}.
\end{eqnarray}

To describe the sensitivity to gravitational waves, we can construct an equivalent energy density spectrum related to these channels\cite{19,24}
\begin{eqnarray}
	\Omega_{A}^{i}(f)=\Omega_{E}^{i}(f)=S_{A}^{i}(f)\frac{4\pi^{2}f^{3}}{3H_{0}^{2}},
\end{eqnarray}
where $H_0$  is the current Hubble constant. For a single Lisa-like detector considering only A and E channels, the total equivalent energy density is\cite{75}
\begin{eqnarray}
	\Omega^i(f)=\frac{4\pi^2f^3}{3H_0^2}\left(\sum_{j=\mathrm{A},\mathrm{E}}\frac{\mathcal{R}_j^\mathrm{i}(f)}{N_j^i(f)}\right)^{-1}.
\end{eqnarray}
\subsection{Sensitivity for Lisa-Taiji networks}
To calculate the energy density for Lisa-Taiji network, we only consider the mutually orthogonal A and E channels, similar to the computation of a single Lisa-like detector. The equivalent energy density formula for the Lisa-Taiji network is given by\cite{73}
\begin{equation}
	\Omega^{cross}(f)=\frac{4\pi^2f^3}{3H_0^2}{\left(\sum_{a=A,E,D=A^{\prime},E^{\prime}}\frac{\mid\gamma_{ab}(f)\mid^2}{S_a^{\mathrm{Lisa}}(f)S_b^{\mathrm{Talji}}(f)}\right)^{-\frac12}},
\end{equation}
here, $S_a^{Lisa}(f)$,$S_b^{Taiji}$ are the noise power spectral density of a single detector, which are given by Eq. \eqref{eq24}, $H_0$ is the current value of the Hubble parameter, and $\gamma_{ab}(f)$ is the overlap reduction function between two different channels of the two triangular detectors. The expression for $\gamma_{ab}(f)$ for the Lisa-Taiji network can be obtained using the ground-based laser interferometer network\cite{78}
\begin{eqnarray}
	\gamma_{ab}=\Theta_{1}(y,\beta)\cos(4\delta)+\Theta_{2}(y,\beta)\cos(4\Delta),
\end{eqnarray}
where
\begin{eqnarray}
	\Delta\equiv\frac{\sigma_1+\sigma_2}2,\delta\equiv\frac{\sigma_1-\sigma_2}2,
\end{eqnarray}
$\sigma_1$,$\sigma_2$are the angles between the bisector of the L-shaped interferometer on each detector and the tangent to the great circle linking the two detectors, calculated counterclockwise. The specific orbit and interferometer positions can be found in references\cite{20,24,73,78}. The function $\Theta_{1}(y,\beta)$ and $\Theta_{2}(y,\beta)$ are defined as
\begin{eqnarray}
	\Theta_1(y,\beta)&=\left(j_0(y)+\frac{5}{7}j_2(y)+\frac{3}{112}j_4(y)\right)\cos^4\left(\frac{\beta}{2}\right),
\end{eqnarray}
\begin{widetext}
	\begin{eqnarray}
		\Theta_2(y,\beta)=&&\left(-\frac38j_0(y)+\frac{45}{56}j_2(y)-\frac{169}{896}j_4(y)\right)+\left(\frac12j_0(y)-\frac57j_2(y)-\frac{27}{224}j_4(y)\right)\cos\beta\nonumber\\
		&&+\left(-\frac18j_0(y)-\frac5{56}j_2(y)-\frac3{896}j_4(y)\right)\cos2\beta,
	\end{eqnarray}
\end{widetext}
where $j_n$  is the $nth$ order spherical Bessel function, $\beta$ is the angle between the information planes of the two detectors, which can be obtained directly by computing their normal vectors. The detector normal vectors can be found in references\cite{31,73}, and $y=2\pi df/c$ is a parameter of the spherical Bessel function, where $d$ is the distance between the two detectors.  For the Lisa-Taiji network, due to the mirror symmetry, $\gamma_{AE^{'}}$=$\gamma_{AE^{'}}$=$0$, so we only need to calculate $\gamma_{AA^{'}}$ and $\gamma_{EE^{'}}$. for the three different Taiji orbit designs. The \textit{m,p,c} orbit model mentioned in paper\cite{24} is used for the design of Taiji's orbit in the joint network. The parameter values related to the overlap reduction function for the three different designs are shown in Table~\ref{table2}.
\begin{table*}
	\caption{\label{table2}Parameter values of the overlap reduction function in three different detector networks.}
	\begin{ruledtabular}
		\begin{tabular}{cccccc}
			\multicolumn{2}{c}{Lisa-Taijip}&\multicolumn{2}{c}{Lisa-Taijim}&\multicolumn{2}{c}{Lisa-Taijic}\\
			$\gamma_{AA^{'}}$&$\gamma_{EE^{'}}$&$\gamma_{AA^{'}}$&$\gamma_{EE^{'}}$
			&$\gamma_{AA^{'}}$&$\gamma_{EE^{'}}$\\ \hline
			$d=1.0\times10^{11}m$&$d=1.0\times10^{11}m$&$d=1.0\times10^{11}m$&$d=1.0\times10^{11}m$&$d=0m$&$d=0m$\\
			$\beta=34.46\si{\degree}$&$\beta=34.46\si{\degree}$&$\beta=71.06\si{\degree}$&$\beta=71.06\si{\degree}$&$\beta=0\si{\degree}$&$\beta=0\si{\degree}$\\
			$\delta=0\si{\degree}$&$\delta=0\si{\degree}$&$\delta=0\si{\degree}$&$\delta=0\si{\degree}$&$\delta=0\si{\degree}$&$\delta=0\si{\degree}$\\
			$\Delta=45\si{\degree}$&$\Delta=0\si{\degree}$&$\Delta=45\si{\degree}$&$\Delta=0\si{\degree}$&$\Delta=45\si{\degree}$&$\Delta=0\si{\degree}$
		\end{tabular}
	\end{ruledtabular}
\end{table*}

\begin{figure}
	\includegraphics[width=\columnwidth]{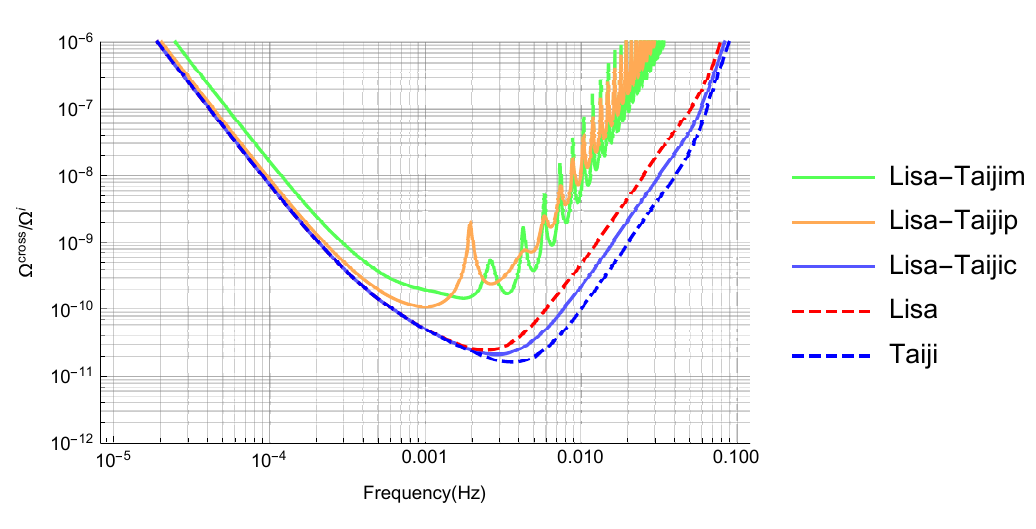}
	\caption{\label{figure2}Sensitivity curves for three different Lisa-Taiji networks and a single detector Lisa, Taiji.}
\end{figure}

The sensitivity curves for the three different networks and the sensitivity curve for a single detector are summarized in Figure\ref{figure2}. It can be found that Lisa-Taijic network has the optimal sensitivity curve among the three different network models. To demonstrate the detection capability of a gravitational wave detector for a power-law random gravitational wave signal with a form similar to $\Omega_h=\Omega_i(f/f_{ref})^{\alpha_i}$, a power-law integrated sensitivity (PLS) was proposed\cite{79}. Based on a given observation time $T_{ob}$ and signal-to-noise ratio (SNR) threshold $\rho_m$,the PLS of the detector is given by\cite{24,31,79}
\begin{eqnarray}
	\Omega_\kappa=\frac{\rho_m}{\sqrt{2T_{ob}}}\bigg(\int_{fmin}^{f_{max}}df\frac{(f/f_{ref})^{2\kappa}}{\Omega_{missions}(f)^2}\bigg)^{-\frac{1}{2}},
\end{eqnarray}
\begin{eqnarray}
	\Omega_{PLS}(f)=\max_\kappa\Omega_\kappa\left(\frac{f}{f_{ref}}\right)^\kappa,
\end{eqnarray}
where the subscript "missions" denotes the joint detector network "cross" or a single Lisa-like detector, $f_{ref}$ can be freely chosen without affecting the PLS result\cite{79}, and the index $\kappa \in[-8,8]$. Based on previous studies of SNR for gravitational wave detectors\cite{31}, we assume $\rho_m=10$ and $T_{ob}=4years$.

\begin{figure}
	\includegraphics[width=\columnwidth]{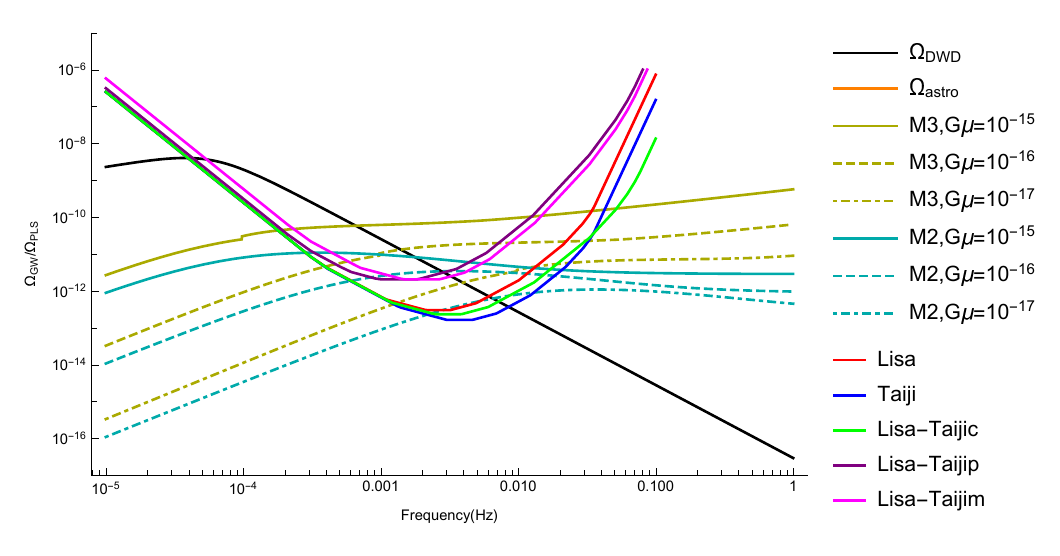}
	\caption{\label{figure3}Gravitational wave from cosmic strings with different parameters and different models, foreground noise of gravitational wave, and PLS plots from different GW detectors for the case $\rho_m=10$ and $T_{ob}=4years$.}
\end{figure}

We integrate the PLS plots for a single Lisa-like detector, the three different joint network configurations, as well as GWs from cosmic strings, DWD, and inspiralling BBHs/BNS in Figure~\ref{figure3}. It can be seen that for Lisa, Taiji, and Lisa-Taiji networks, they can all detect the cosmic string signal in Model 2 with $G\mu=10^{-17}$ (since Model 1 and Model 2 can be expressed by the same formula\cite{30,32}, we use Model 2 for simplicity in the following discussion). However, the sensitivity of Lisa-Taijic is better than Lisa, while Taiji has better sensitivity than Lisa-Taijic in the range of $1\sim20$mHz, and the sensitivity of Lisa-Taijic in other frequency ranges is superior to all of the above detectors and other detector networks.

\section{\label{IV}ESTIMATION AND MODELING}
It can be concluded from Figure~\ref{figure2} and Figure~\ref{figure3} that Lisa-Taijic is the optimal choice in the joint networks. Lisa, Taiji and Lisa-Taijic are capable for detecting SGWB with cosmic string tensions $G\mu>10^{-17}$ under all the cosmic string models considered in this paper. The sensitivity of Lisa-Taijip and Lisa-Taijim is significantly weaker than that of Lisa-Taijic, Lisa and Taiji. Therefore, when performing parameter estimation, we only consider the Lisa-Taijic network for the joint detection network, and we only consider the A and E channels for TDI.

In this section, we adopt Fisher Information Matrix to calculate the ability of Lisa, Taiji and Lisa-Taijic for estimating the cosmic string tension $G\mu$ in different data cases, and the DIC method is used to calculate when the cosmic string tension $G\mu$ reaches the point where Lisa, Taiji and Lisa-Taijic can provide evidence of a cosmic string signal in the detection data if the detection data contain confusion noise.

\subsection{The Fisher Information Matrix}
We considered three data cases for the data $d^i$ in a single GW detector:\\
1) $d^i=\Omega_{A,E}^i(f)+\Omega_{GW}(f,M_x)$, with parameter space $\theta^i=\{N_a^i,N_o^i,G\mu\}$, i.e., detector noise mixed with GW signals generated by different cosmic strings.\\
2) $d^i=\Omega_{A,E}^i(f)+\Omega_{astro}(f)+\Omega_{DWD}(f)$, with parameter space $\theta^i=\{N_a^i,N_o^i,A_1,\alpha_1,A_2,\alpha_2,\Omega_{astro},\alpha_{astro}\}$, i.e., detector noise, double white dwarf foreground noise, and gravitational wave background of inspiralling BBHs/BNS.\\
3) $d^i=\Omega_{A,E}^i(f)+\Omega_{astro}(f)+\Omega_{DWD}(f)+\Omega_{GW}(f,M_x)$, with parameter space $\theta^i=\{N_a^i,N_o^i,A_1,\alpha_1,A_2,\alpha_2,\Omega_{astro},\alpha_{astro},G\mu\}$, i.e., detector noise, double white dwarf foreground noise, gravitational wave background of inspiralling BBHs/BNS , and SGWB generated by cosmic strings.

For Lisa-Taijic, we also consider the above three data and the corresponding parameter space is as follows:\\
	1) $\theta=\{N_a^{Lisa},N_o^{Lisa},N_a^{Taiji},N_o^{Taiji},G\mu\}$,\\
	2) $\theta=\{N_{\alpha}^{Lisa},N_{o}^{Lisa},N_{\alpha}^{Taiji},N_{o}^{Taiji},A_{1},\alpha_{1},A_{2},\alpha_{2},\\
	~~~~~~~~~~~~\Omega_{astro},\alpha_{astro}\}$,\\
	3) $\theta=\{N_a^{Lisa},N_o^{Lisa},N_a^{Taiji},N_o^{Taiji},A_1,\alpha_1,A_2,\alpha_2,\\
	~~~~~~~~~~~~\Omega_{astro},\alpha_{astro},G\mu\}.$
	
The likelihood function for a single Lisa-like detector can be constructed from the frequency domain data $(d^i=\{d_A^i,d_E^i\})$ and the given model parameters $\theta$\cite{19,54,80}
\begin{equation}
	\mathcal{L}^i(d\mid\theta^i)=\prod_{\alpha=0}^N\frac1{\sqrt{\det(2\pi\mathcal{C}^i(\theta^i,f_\alpha))}}e^{-\frac12{d_\alpha^i}^{*T}{\mathcal{C}^i}^{-1}(\theta,f_\alpha)d_\alpha^i}\label{34}.
\end{equation}

For the joint network, the likelihood function has the following form
\begin{equation}
	\mathcal{L}\left(d\mid\theta\right)=\prod_{\alpha=0}^N\frac1{\sqrt{\det(2\pi\mathcal{C}\left(\theta,f_\alpha\right))}}e^{-\frac{1}{2}d_{\alpha}^{l}{}^{*T}\mathcal{C}^{-1}(\theta,f_{\alpha})d_{\alpha}^{r}}\label{35},
\end{equation}
where $\alpha$ represents the frequency point. In the joint network
\begin{eqnarray}
	d_\alpha^l=\{d_A^{lisa},d_E^{lisa},d_A^{Taiji},d_E^{Taiji},d_A^{lisa},d_E^{lisa}\},
\end{eqnarray}
and
\begin{eqnarray}
	d_\alpha^r=\{d_A^{lisa},d_E^{lisa},d_A^{Taiji},d_E^{Taiji},d_A^{Taiji},d_E^{Taiji}\}.
\end{eqnarray}

The data composition and parameter space for different data case can be found in the first and second paragraphs of this section. $\mathcal{C}^i(\theta^i,f_\alpha)$ is the power spectral covariance matrix of a single Lisa-like detector at frequency point $\alpha$. Its form is
\begin{equation}
	\mathcal{C}^i(\theta^i,f_\alpha)=\begin{pmatrix}S_{h,A}^i(f_\alpha)+N_A^i(f_\alpha)&0\\0&S_{h,E}^i(f_\alpha)+N_E^i(f_\alpha)\end{pmatrix}.
\end{equation}

For simplicity, we omit $f_\alpha$ in the covariance matrix of the joint detector and write Lisa-Taijic as $LTc$, which is
\begin{equation}
	\mathcal{C}\left(\theta,f_\alpha\right)=\begin{pmatrix}\mathcal{C}^{lisa}(\theta^{lisa},f_\alpha)&0&0\\0&\mathcal{C}^{Taiji}(\theta^{Taiji},f_\alpha)&0\\0&0&\mathcal{C}^{LTc}(\theta,f_\alpha)\end{pmatrix}.\label{eq39}
\end{equation}

$N_I^i(f_\alpha)$ is the noise power spectral density of different TDI channels, where $I={A,E}$. The power spectral density of signals in different channels of a single detector is given by
\begin{eqnarray}
	S_{h,l}^i(f_\alpha)=\frac{3H_0^2\Omega_{\mathrm{tot}}(f_\alpha)}{4\pi^2{f_\alpha}^3}\mathcal{R}_l^\mathrm{i}(f_\alpha),	
\end{eqnarray}
the signal power spectral density for the joint detector is
\begin{eqnarray}
	S_{h,II^{\prime}}^{i}(f_{\alpha})=\frac{3H_{0}^{2}\Omega_{\mathrm{tot}}(f_{\alpha})}{4\pi^{2}f_{\alpha}^{3}}\gamma_{II^{\prime}}(f_{\alpha}),
\end{eqnarray}
here, $\Omega_{tot}$ is the total energy density of the GW in the data case. Due to mirror symmetry, $\gamma_{AE^\prime}=$$\gamma_{EA^\prime}$=0, Eq.~\eqref{eq39} can be expanded as
\begin{widetext}
	\begin{eqnarray}
		\mathcal{C}\left(\theta,f_\alpha\right)=diag&(S_{h,A}^{lisa}+N_A^{lisa},S_{h,E}^{lisa}+N_E^{lisa},S_{h,A}^{Taiji}+N_A^{Taiji},S_{h,E}^{lisa}+N_E^{lisa},\nonumber\\&S_{h,AA^{\prime}}+\sqrt{N_A^{lisa}*N_A^{Taiji}},S_{h,EE}+\sqrt{N_E^{lisa}*N_E^{Taiji}}).
	\end{eqnarray}
\end{widetext}

\begin{figure}
	\includegraphics[width=\columnwidth]{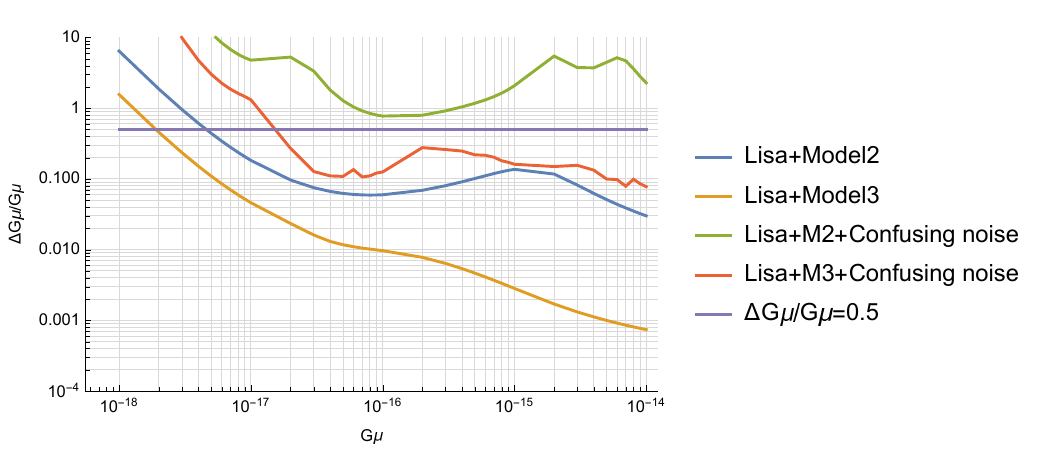}
	\caption{\label{figure4}Fisher matrix estimation of the uncertainty of cosmic string tension $G\mu$ for Lisa detector under different models. The purple horizontal line represents uncertainty $\Delta G\mu/G\mu=0.5$. The blue and orange solid lines represent data case 1); the green and red solid lines represent data case 2).}
\end{figure}

\begin{figure}
	\includegraphics[width=\columnwidth]{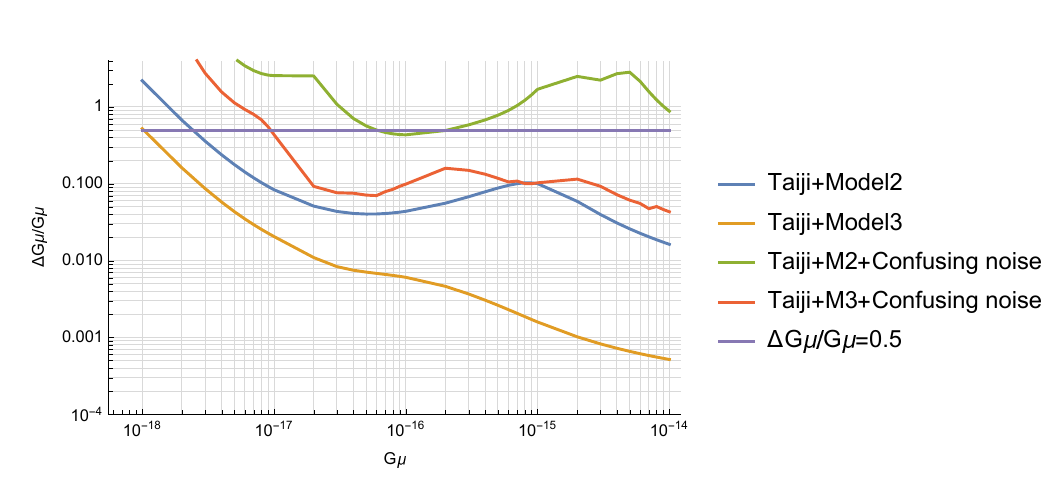}
	\caption{\label{figure5}Fisher matrix estimation of the uncertainty of cosmic string tension $G\mu$ for Taiji detector under different models. The purple horizontal line represents uncertainty $\Delta G\mu/G\mu=0.5$. The blue and orange solid lines represent data case 1); the green and red solid lines represent data case 2).}
\end{figure}

The Fisher information matrix (FIM) is commonly used to estimate the uncertainty of gravitational wave parameters\cite{24,54,75,81,82,83}. We use data cases 1) and 3) as observation data to calculate the Fisher matrix. For a parameter a in the parameter space of a specific data, its uncertainty is expressed by the standard deviation of that parameter $\sqrt{F_{aa}^{-1}}$.  The Fisher matrix can be constructed from the covariance matrix\cite{18,19}, and the result is similar to the literature\cite{24,54,75,81}. For a single detector, the uncertainty estimation of the cosmic string tension $G\mu$ using the Fisher matrix is shown in Figure~\ref{figure4} and Figure~\ref{figure5}. The form of the Fisher matrix is given by
\begin{equation}
	F_{ab}^i=2\sum_{I=A,E}T_{\mathrm{ob}}\int_0^{f_{max}}\frac{\partial\ln\mathcal{C}^i{}_{II}(f_\alpha)}{\partial\theta_a}\frac{\partial\ln\mathcal{C}^i{}_{II}(f_\alpha)}{\partial\theta_b}df_\alpha.
\end{equation}

\begin{figure}
	\includegraphics[width=\columnwidth]{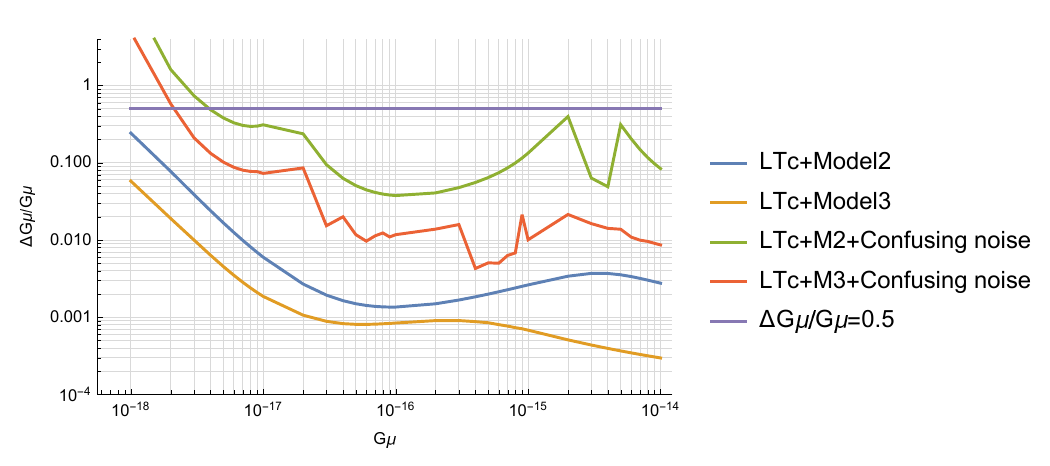}
	\caption{\label{figure6}Fisher matrix estimation of the uncertainty of cosmic string tension $G\mu$ for Lisa-Taiji network under different models. The purple horizontal line represents uncertainty $\Delta G\mu/G\mu=0.5$. The blue and orange solid lines represent data case 1); the green and red solid lines represent data case 2).}
\end{figure}

For the $LTc$ network, the Fisher matrix for estimating the uncertainty of the cosmic string tension $G\mu$ is shown in Figure~\ref{figure6}, and its form is
\begin{equation}
	F_{ab}=2\sum_{l=1}^6T_{\mathrm{ob}}\int_0^{f_{max}}\frac{\partial\ln\mathcal{C}_{ll}(f_\alpha)}{\partial\theta_a}\frac{\partial\ln\mathcal{C}_{ll}(f_\alpha)}{\partial\theta_b}df_\alpha.
\end{equation}

Neglect the detector noise and assume that the signal $S_{h}\ll N,$,then the FIM of $LTc$ network can be simplified to the results in literature\cite{24,81}. When the SGWB of cosmic strings is in $M_3$ form, we can make the most accurate estimate of the cosmic string tension $G\mu$ through analysing the detector data. This means that if the SGWB from cosmic strings follows $M_3$, the string tension $G\mu$ can be estimated accurately. Even when considering data of $M_3$, which includes foreground noise from DWD and inspiralling BBHs/BNS, the detector's constraint on the string tension in $M_3$ will still be better than that in $M_2$ by an order of magnitude. We use the same level line $\Delta G\mu/G\mu=0.5$ as in literature\cite{18} to illustrate the estimation capability of a single detector and $LTc$ network for cosmic string tension under different observation scenarios. 

\begin{figure}
	\includegraphics[width=\columnwidth]{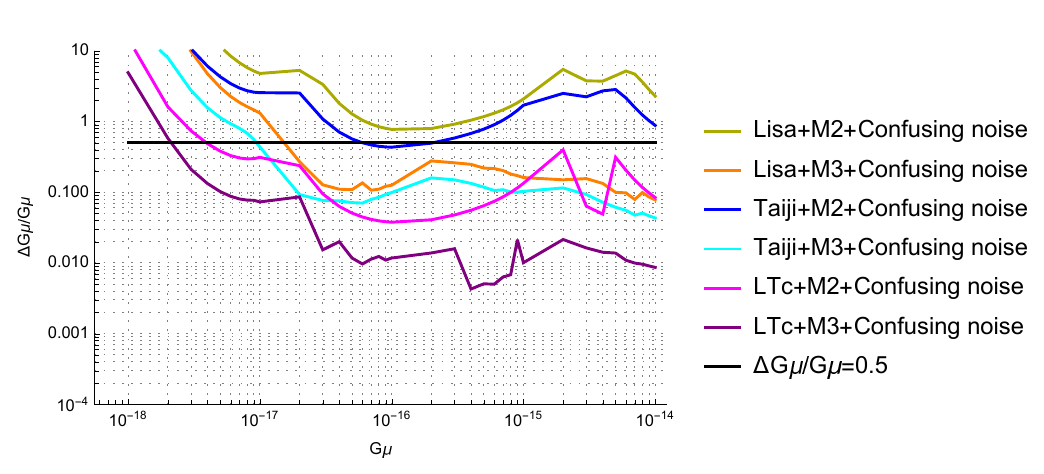}
	\caption{\label{figure7}Fisher matrix estimation of the uncertainty of cosmic string tension $G\mu$ for gravitational wave detectors under different models in data case 3). The dark green and orange solid lines represent Lisa; the blue and teal solid lines represent Taiji; the pink and purple solid lines represent Lisa-Taijic network; the horizontal black solid line represents uncertainty $\Delta G\mu/G\mu=0.5$.}
\end{figure}

\begin{figure}
	\includegraphics[width=\columnwidth]{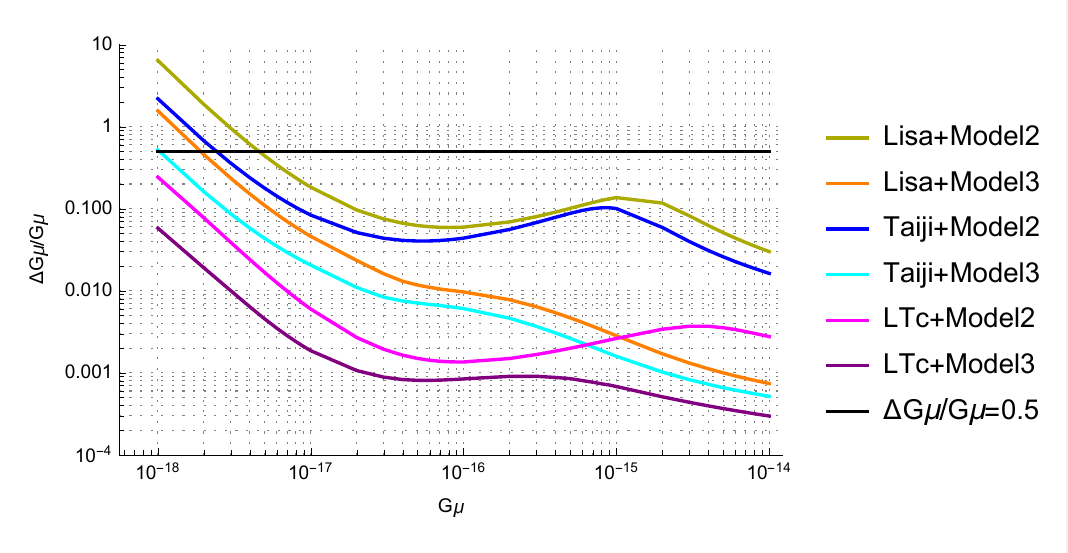}
	\caption{\label{figure8} Fisher matrix estimation of the uncertainty of cosmic string tension $G\mu$ for gravitational wave detectors under different models in data case 1). The dark green and orange solid lines represent Lisa; the blue and teal solid lines represent Taiji; the pink and purple solid lines represent Lisa-Taijic network; the horizontal black solid line represents uncertainty $\Delta G\mu/G\mu=0.5$.}
\end{figure}

It can be seen that for a single detector Taiji has a better restriction ability for cosmic string tension than Lisa. The estimation of relative uncertainty on cosmic string tension in Lisa-Taiji network is significantly better than that in a single detector. To compare the uncertainty estimation of cosmic string tension $G\mu$ for different data case between the single detectors and joint network, we show the results of detector under data case 3) in Figure~\ref{figure7} and data case 1) in Figure~\ref{figure8}. The results show that Lisa-Taijic network has a better restriction ability for cosmic string tension $G\mu$ than a single detector in any model and data case, and its uncertainty estimation ability for cosmic string tension $G\mu\in\{10^{-17}\sim10^{-16}\}$ can be improved by about one order of magnitude.
\subsection{Deviance Information Criterion}
We compare the data produced by a single detector as well as the $LTc$ network for data case 1) and data case 3) to investigate the detectability of SGWB from cosmic strings in the presence of DWD foreground and a background of inspiralling BBHs/BNS, as well as the accuracy of the estimation of the cosmic string tension $G\mu$. In order to investigate whether the detector provides a better fit to a data case that includes cosmic strings or does not, as well as to the detectability of SGWB from cosmic strings, we use the deviance information criterion (DIC) for model comparison, which can be used even with inappropriate or vague priors\cite{18,19,84,85}. The calculation of DIC requires Markov chain Monte Carlo (MCMC) sampling firstly, where the variance of the posterior samples $D(\theta)$  and the penalty term $p_D=\bar{D}-D(\bar{\theta})$ are calculated after MCMC sampling, then the Bayesian factor DIC is obtained as
\begin{eqnarray}
	DIC=D(\bar{\theta})+2p_D.
\end{eqnarray}

Where $\theta$ is the posterior sample mean of the parameter $\theta$, $D(\theta)$ is defined as $D(\theta)=-2\log\mathcal{L}(d\mid\theta),$, and $\bar{D}$ is the posterior mean of the variance. In calculating DIC, data case 2) and 3) were used as observed results. Whether the detector can provide evidence for data containing cosmic strings can be determined by calculating the difference in DIC between the  DIC of detector for the case of data with cosmic strings and the case of data without cosmic strings\cite{18,86,87}. An adaptive Markov chain Monte Carlo\cite{88} was used for sampling, based on the Metropolis-Hastings algorithm. For MCMC, a prior distribution and a posterior distribution constructed from a likelihood function are needed. The likelihood functions for different detectors are given by Eq.~\ref{34} and Eq.~\ref{35}, and the prior distribution is assumed to be an independent Gaussian distribution as shown
\begin{eqnarray}
	p(\theta)=\prod_n\operatorname{Exp}\left(-\frac{(\theta_n-\mu_n)^2}{2\sigma_n^2}\right),\label{4.14}
\end{eqnarray}
where for data case 3), $\mu_n$ represents the true values of detector noise, DWD foregrounds, background of inspiralling BBHs/BNS, and cosmic string parameters, and $\sigma_n$  is the variance assume $\sigma_n =1$. Similar to reference\cite{19}, logarithmic parameter sampling was used for $N_a^i$, $N_o^i$, $A_1\text{,}A_2\text{,}\Omega_{astro},G\mu.$ While direct sampling was used for $\alpha_1$, $\alpha_2$, $\alpha_{astro}$. In the detectable frequency range, the likelihood function was constructed by equally dividing each unit logarithmic frequency range into ten parts. Therefore, the posterior distribution of the joint $LTc$ network for data case 3) can be obtained by combining Eq.~\ref{35} and Eq.~\ref{4.14} as shown
\begin{eqnarray}
	p(\theta|d)\propto p(\theta)\mathcal{L}(d\mid\theta).
\end{eqnarray}

An adaptive Metropolis-Hastings algorithm\cite{88} was used in MCMC sampling to improve acceptance rates by using a proposal distribution $Q_m(\theta)$, and the proposed distribution for the $mth$ iteration is in the form
\begin{equation}
	Q_m(\theta)=(1-\beta)\mathcal{N}(\theta,(2.28)^2\Sigma_m/d)+\beta\mathcal{N}(\theta,(0.1)^2\mathcal{I}_d/d),
\end{equation}
where $\beta=0.01$, $\mathcal{N}$ is a multivariate normal distribution, $\Sigma_m$ is the current empirical estimation of the covariance matrix of the parameter vector $\theta$ at the $mth$ iteration, $d$ is the number of parameters, $\mathcal{I}_d$ is a $d$-dimensional identity matrix. The number of parameters varies depending on the data case and detector.

MCMC sampling is performed for different detectors under data case 2) and 3) with a sampling iteration of $m=200000$, and the covariance matrix is estimated empirically based on 2000 samples. Since there is randomness in sampling, there is also randomness in DIC results. To reduce the impact of randomness, we perform ten separate samplings with different cosmic string tensions $G\mu$ for each data case and detector, and calculate the DIC value using the posterior samples from ten separate samplings, and take their average as the final result.
\begin{figure}
	\includegraphics[width=\columnwidth]{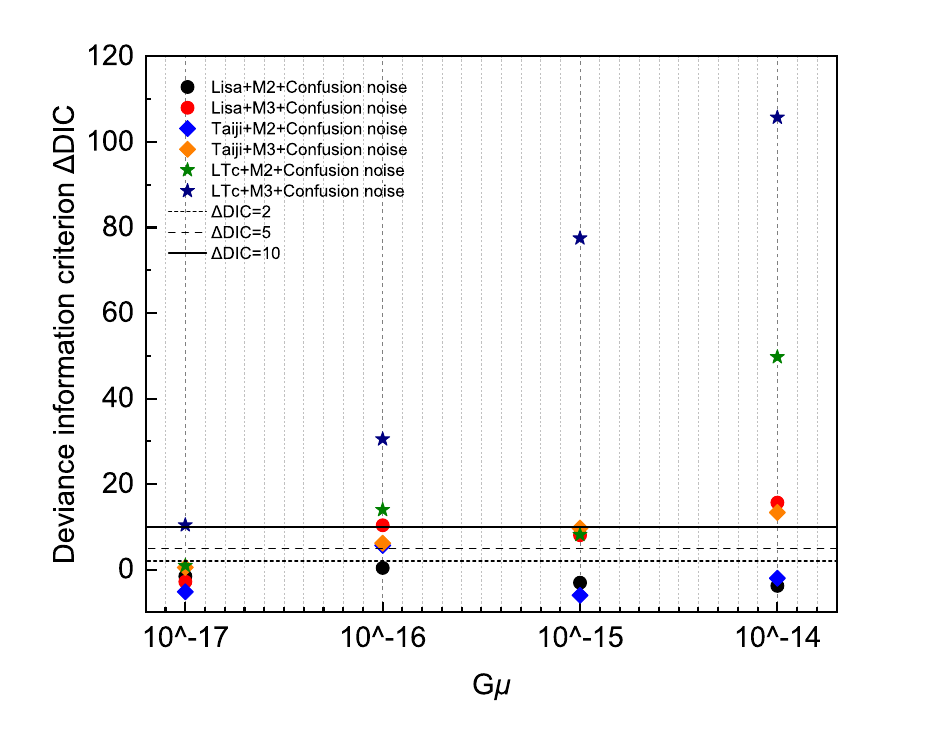}
	\caption{\label{figure9}Model comparison of different detectors using DIC for data case 2) and 3). The pentagram represents the DIC difference of the  $LTc$ network for different data cases, the circle represents Lisa, and the diamond represents Taiji}
\end{figure}

The DIC values for different detectors under data case 3) and the difference in DIC values for different detectors under data case 2) are shown in Figure 9. Following the general empirical rule, when $\Delta DIC>2$, evidence for data case with cosmic strings begins to be provided, when $\Delta DIC>5$, there is sufficient evidence to prove the presence of cosmic strings in the data case, and when $\Delta DIC>10$, there is strong and decisive evidence\cite{18,19,86,87}. It can be seen that the DIC results show a similar trend as the FIM results, and $LTc$ network has significantly improved detectability compared to a single detector for SGWB in both cosmic string models. For the cosmic string model in $M_3$, all detectors provide sufficient evidence for the presence of SGWB from cosmic string in the data case with $G\mu\sim10^{-16}$ and $G\mu>10^{-16}$, while the $LTc$ network provides decisive evidence with $G\mu\sim10^{-17}$ and $G\mu>10^{-17}$. For the cosmic string model $M_2$, only the $LTc$ network provides evidence from $G\mu\sim10^{-17}$, while a single detector can only provide evidence at $G\mu\sim10^{-16}$ and cannot provide sufficient evidence at other positions.
\section{\label{V}CONCLUSION}
We mainly focus on the constraints on the cosmic string tension $G\mu$ in different observation data cases using a single mHz detector and the joint networks. We specifically calculate the detectability of SGWB from cosmic string in different models and different data cases. We compared the equivalent energy density curve and PLS curve of a single detector and three different Taiji orbits combined with Lisa. The SGWB from cosmic string with tension $G\mu=10^{-17}$ can be detected by a single detector and the $LTc$ network. 

Furthermore, we calculate the uncertainty in the parameter estimation of the cosmic string tension when the observation data case is a combination of SGWB from cosmic string and the foreground noise. We use the Fisher information matrix for parameter estimation and the DIC method for detectability analysis. The results suggest that the $LTc$ network has better performance than a single detector in terms of parameter estimation and detectability of cosmic strings. According to the results from Fisher information matrix, for observation data case with foreground noise, the $LTc$ network shows the best performance in different cosmic string models. The uncertainty of cosmic string tension is $\Delta G\mu/G\mu<0.5$ since $G\mu\sim4\times10^{-17}$. For data case only containing SGWB from cosmic strings, it can achieve  $\Delta G\mu/G\mu<0.5$ in lower tension regions. According to the DIC results, the $LTc$ network also exhibits better properties than a single detector. It provides evidence for the existence of $M_3$ from $G\mu\sim10^{-17}$ and evidence for $M_2$ from $G\mu\sim10^{-16}$. Therefore, using a joint Lisa-Taiji network to observe SGWB from cosmic string may be a good choice in practical applications. 

In this paper, we only consider foreground including modulated DWD and GWB model generated by BBHs/BNS from on LIGO and Virgo. In actual observations, more confusion gravitational waves need to be considered. For the cosmological SGWB, we only consider the one from cosmic strings. However, in reality, there will be more scientific requirements, such as searching for cosmological stochastic gravitational wave generated by first-order phase transitions\cite{89} or inflation\cite{90}. Therefore, our next step is to consider the estimation of parameters related to first-order phase transitions using joint detectors. At the same time, we are aware of another space-based detector called TianQin\cite{91}, which plays a unique role in the high-frequency region through its own joint observations or joint observations with Lisa and Taiji\cite{73}. In our subsequent studies, we will also consider TianQin detectors for joint observations to search for cosmological stochastic gravitational wave.

\begin{acknowledgments}
	This work was supported by the National Key Research and Development Program of China (Grant No. 2021YFC2203004), the National Natural Science Foundation of China (Grant No. 12147102), the Natural Science Foundation of Chongqing (Grant No. CSTB2023NSCQ-MSX0103).
\end{acknowledgments}

\nocite{*}

\bibliography{apssamp}

\end{document}